\documentstyle[preprint,aps]{revtex}
\begin{document}

\title{
Quasiparticle Spectra in the Hubbard Model.
}
\author
{William H. Beere and James F. Annett}

\address{University of Bristol,
H.H Wills Physics Laboratory,\\
Royal Fort, Tyndall Avenue, Bristol BS8 1TL, UK}
\maketitle

\begin{abstract}
 We examine the quasiparticle lifetime and spectral weight
near the Fermi surface in the two-dimensional Hubbard model.
We use the FLEX approximation to self-consistently generate the
 Matsubara Green's functions 
and then we analytically continue to the real
axis to obtain the quasiparticle
 spectral functions. 
We compare the spectral functions
found  in the nearest neighbor hopping only
Hubbard model with those  
found when the second  neighbor hopping is included.
This separates the effects
of nesting, the van Hove singularity and short-ranged
antiferromagnetic correlations.  The quasiparticle
scattering rate is enhanced along the
$(0,\pi)$ to $(\pi,0)$ Brillouin zone diagonal.
When the density is  close to half-filling these `hot spots' lie on
the Fermi surface and the scattering rate increases with decreasing
temperature.
For the next-nearest neighbor hopping scenario we observe a large range
of doping where there is no antiferromagnetism but the scattering
rate has a linear temperature dependence.  On decreasing the interaction
this non-Fermi liquid behavior is confined to doping 
levels where the Fermi energy lies near the Van Hove singularity.
We conclude that the `hot spots' are associated with the antiferromagnetic
phase transition while the linear temperature dependence of the scattering
rate is associated with the Van Hove singularity.
\end{abstract}

\section{Introduction}
The Hubbard model\cite{hubbard}
is one of the simplest
models for electronic correlations in solids and yet it is  still not fully
understood. 
If the electron-electron interaction strength, $U$, is positive
and small compared to the electronic band-width, $W$, the model is 
generally expected to
have a normal Fermi liquid ground state
in two or three dimensions\cite{brinkman}.
At higher $U/W$ a number of different phases may exist, including: 
an antiferromagnetic insulator state, ferromagnetism, and
d-wave superconductivity\cite{hirsch,gross}.
It was  also proposed that in two-dimensions 
metallic non-Fermi liquid states may exist at
large $U/W$, such as the 
resonant-valence-bond (RVB)\cite{baskaran} or Luttinger 
liquid\cite{anderson} state, and the flux-phase\cite{affleck}.
Whether or not these phases actually exist as stable ground states
of the Hubbard model
remains controversial, despite considerable numerical and analytic
effort to solve the two dimensional Hubbard model. Similarly, whether or
not d-wave superconductivity occurs in the positive $U$ Hubbard
model is also unclear at the present time. 
Of course both non-Fermi liquid behavior and d-wave superconductivity
appear to occur experimentally in the cuprates\cite{varma,annett},
where it is generally believed that the Hubbard model is appropriate.
Indeed the two-dimensional Hubbard model can be
derived for the cuprates from the \textit{ab initio} density functional 
electronic structure
under certain simplifying assumptions\cite{bacci}. However it still remains
unclear whether the Hubbard model is sufficient to explain all of
 the anomalous
properties of the cuprates, or whether one must also
include physical effects beyond the Hubbard Hamiltonian,
such as electron-phonon coupling or multiple electronic bands.

In this paper we use the fluctuation exchange approximation\cite{bickers},
FLEX, to
investigate the electronic structure and phase diagram for the 
two-dimensional Hubbard model.
We pay particular attention to the antiferromagnetic (AFM) phase and the 
effect of antiferromagnetic fluctuations near to the AFM transition.
The theory of nearly-antiferromagnetic Fermi 
 liquids has been recently developed by Pines and
co-workers\cite{pines}. Although based upon a Fermi liquid approach, the
theory explains some of the non-Fermi liquid like properties of the
cuprates in terms of 
the appearance of `hot spots' on the Fermi surface where the 
quasiparticle scattering rate is unusually high.
Altmann, Brenig and Kampf\cite{kampf} have recently shown that the 
FLEX approximation is able to describe these `hot spots'. Below we
investigate in detail  the regions of parameter space 
(e.g. temperature, $T$, band filling, $n$, and hopping $t'/t$) 
 where these `hot spots' occur.

A second experimental feature of the cuprates is the appearance of
a temperature scale $T^*$ (or more than one such temperature scale) distinct
from both the N\'eel temperature $T_N$ and the superconducting transition
temperature $T_c$.  Below $T^*$ a loss of spectral weight occurs
in the low energy excitations, referred to as a pseudogap or 
spin-gap\cite{timusk,hayden}.
The pseudogap can be seen directly in the angle resolved photoemission
experiments\cite{ding} (ARPES) as a small k-dependent shift of the
 spectral edge 
relative to an absolute reference such as a good metal in electrical
contact
with the superconductor.  Below we discuss the appearance of
both $T^*$ and the
pseudogap in terms of the FLEX calculations for the Hubbard model.

Finally we also examine in detail the effects of the Van Hove singularity
in the density of states in the two dimensional Hubbard model. 
The existence of van Hove singularities close to the Fermi energy
is a feature of {\it ab initio} band structures of the 
cuprates\cite{okanderson}. 
The presence of the Van Hove
singularity near to the Fermi surface can lead to a dramatic
increase in $T_c$ as a function of doping\cite{newns,gyorffy}.
The proximity to the Van Hove singularity can also be a source of 
non-Fermi liquid electron-electron scattering rates.
We show below that the `hot spots' and the $T^*$ behavior are assosciated
with the antiferromagnetism, while the approximately linear in $T$ scattering 
rate derives from the proximity to the Van Hove singularity.
At small $U$ (and $t'\neq 0$) these two effects exist quite separately at 
different band fillings, while at larger $U$ (or $t'=0$) both
effects coexist for a wide range of doping levels.

\section{The Fluctuation Exchange Approximation}

The FLEX method was originally introduced by Bickers and
 Scalapino\cite{bickers}. It has been
discussed by many previous authors\cite{BSW,DHS,BW,Yone}, so we
present only a brief outline.
FLEX makes use of the similarity of the `ladder' diagrams to the
`bubble' diagrams from RPA when the electron-electron interaction
 is an on-site interaction, as in the Hubbard model:
\begin{equation}
  \hat{H} = \sum_{<ij>,\sigma} t_{ij} c^{\dagger}_{i\sigma} c_{j\sigma}
 + U \sum_i n_{i\uparrow} n_{i\downarrow} .
\end{equation}
 FLEX
sums both the ladder and RPA bubble diagrams, it is exact to order
$U^3$ (while RPA is only exact to $O(U^2)$) and is a conserving 
approximation in the sense of Baym\cite{Baym}.

The electron Green's function is given by,
\begin{equation}
G({\mathbf k},i\omega_n ) = {1 \over i\omega_n - \xi_{\mathbf k}
- \Sigma({\mathbf k},i\omega_n ) }, \label{eq:green}
\end{equation}
where $\omega_n =  2 \pi (n + 1/2)  T$ is the Matsubara frequency, T
the temperature,
$\Sigma({\mathbf k},i\omega_n )$ 
the self-energy and $\xi_{\mathbf k}$ is the non-interacting band energy.
For a 2d Hubbard model system with nearest and next-nearest neighbor 
hopping $t_{ij}$ this is given by,
\begin{equation}
\xi_{\mathbf k} = -2  t \left [ \cos{ ( k_x)} + \cos{ ( k_y )}\right ]
 - 4 t'
\cos{(k_x )} \cos{ ( k_y  )} - \mu,
\end{equation}
where $t$ is the nearest neighbor hopping energy,  $t'$ the
next-nearest neighbor hopping energy and $\mu$ is the chemical potential.

For FLEX both the particle-particle and particle-hole pair correlators need
to be calculated, as
\begin{eqnarray}
\chi^{ph} ( {\mathbf q},i\omega_m ) &=& - {1 \over T }
\sum_{\omega_n} \int d^2k\ 
G({\mathbf k}, i\omega_n ) G( {\mathbf {q + k}, i\omega_m + i\omega_n }
)\label{eq:chiph} \\
\chi^{pp} ( {\mathbf q},i\omega_m ) &=&  {1 \over T }
\sum_{\omega_n}  \int d^2k\ 
G( {\mathbf k},i\omega_n  ) G( {\mathbf {q - k}}, i\omega_m - i\omega_n 
). \label{eq:chipp}
\end{eqnarray}
The FLEX approximation for the self-energy is then given by,
\begin{eqnarray}
\Sigma({\mathbf k}, i\omega_n) & = & \sum_{\omega_m} \int d^2q
 \left[ 
G( {\mathbf{ k-q}}, i\omega_n - i\omega_m )
 V^{(2)}( {\mathbf{q}}, i\omega_m) \right. \nonumber \\
& & + G( {\mathbf{ k-q}},i\omega_n - i\omega_m)
 V^{(ph)}({\mathbf{ q},i\omega_m}) \nonumber \\
& &  \left.  + G( {\mathbf{ -k+q}}, -i\omega_n + i\omega_m) V^{(pp)}(
 {\mathbf{ q}, i\omega_m}) \right]. \label{eq:sigma}
\end{eqnarray}
Following Bickers and Scalapino the interaction vertices,
$V^{(2)},V^{(ph)}$ and $V^{(pp)}$ are given by,
\begin{eqnarray}
V^{(2)}( {\mathbf{ q}}, i\omega_m) & = & U^2
 \chi^{ph}({\mathbf{ q}},i\omega_m)\\
V^{(ph)}( {\mathbf{ q}}, i\omega_m) &=& {1\over 2} U^2
\chi^{ph}({\mathbf{ q}},i\omega_m)
\left[
{1\over 1 + U \chi^{ph}( {\mathbf{ q}},i\omega_m) } - 1
\right] \nonumber \\
& & +  {3 \over 2} U^2 \chi^{ph}({\mathbf{ q}},i\omega_m)
\left[
{1\over 1 - U \chi^{ph}( {\mathbf{ q}},i\omega_m) } - 1
\right]\\
V^{(pp)}( {\mathbf{ q}},i\omega_m) &=& - U^2
 \chi^{pp}({\mathbf{ q}},i\omega_m)
\left[
{1\over 1 + U \chi^{pp}( {\mathbf{ q}},i\omega_m) } - 1
\right] . \label{eq:vpp}
\end{eqnarray}
Note that some authors have used a modified version of FLEX in which
the particle-particle channel scattering is suppressed\cite{kampf}.
The close similarity of our numerical results to theirs suggests that
the effects of the particle-particle channel are relatively minor.
The Hartree-Fock terms have been omitted from these equations as the
constant term which they produce has been explicitly incorperated into
the chemical potential.

Using the above set of  equations we have suppressed antiferromagnetic
ordering by including only matrix elements diagonal  in spin space.
Nevertheless we can still calculate the antiferromagnetic ordering
temperature
within FLEX, since the instability of the paramagnetic state is given by a 
Stoner like criterion:
\begin{equation}
U \chi^{ph}( {\mathbf Q}_{AF} , 0  )  > 1. \label{aftc}
\end{equation}
Here the antiferromagnetic ordering wave vector is given by 
 ${\mathbf Q}_{AF}=(\pi,\pi)$,  which is the 
position of the main peak in $\chi^{ph}$.
In all cases we only show results obtained in the `normal' metallic
state computed outside of the 
antiferromagnetic phase, given by Eq. \ref{aftc}. We
assume that our calculations are outside any superconducting
regions in the phase diagram.

The system of  equations Eqs. \ref{eq:green}-\ref{eq:vpp} 
form a self-consistent set.
The Brillouin zone integrals were discretized and we used either
a $32\times 32$ or $64\times 64$ grid depending on the accuracy required.
The Matsubara frequency sums were typically carried out using
1024 points. For temperatures of $0.06t$ this relates to a 
Matsubara frequency cut-off of about $\pm 200t$.
All of the momentum and frequency convolutions in Eqs. \ref{eq:chiph},
 \ref{eq:chipp},
\ref{eq:sigma} were done using fast Fourier transforms.  The frequency
convolution leads to
high frequency errors in the correlation functions and the self-energy,
but this does not produce significant errors in the final 
low frequency part of the Green's function.

In order to obtain the real frequency spectral function
$A({\mathbf k},\omega)$,
\begin{equation}
A({\mathbf k},\omega) = - {1\over \pi} \mbox{Im}
 G({\mathbf k},\omega+i\epsilon). 
\end{equation}
 the Green's functions were
analytically continued from the Matsubara frequencies to the real axis.
To do this we fit the Green's functions to a continued fraction
of the form
\begin{equation}
G(z) = \frac{\textstyle a_1}{\textstyle 1+
 \frac{\textstyle a_2(z-z_1)}{\textstyle 1+
\frac{\textstyle \rule{0mm}{3mm}\dots}{\textstyle 1+a_N(z-z_{N-1})}}}
\end{equation}
where the coefficients $a_n$ are determined from the known Green's
function for the positive Matsubara frequencies $z_1 =  i\omega_0$,
$z_2 =  i\omega_1$ etc.\cite{paderef}.
Typically we used up to 512 terms in the continued fraction.
The continued fraction approximant for $G(z)$ defines the retarded
Green's function in the upper-half plane. 
Evaluating this on the real axis then gives $G(\omega+i\epsilon)$
and hence the spectral function,  $A({\mathbf k},\omega)$.

The chemical potential, $\mu$,  was kept fixed during the self-consistent
runs,
and then the band filling, $\langle n \rangle$,  was calculated from the final 
Green's function
\begin{equation}
  \langle n \rangle = - \sum_{\omega_n} \int d^2k
 \left( \frac{1}{2}+ G({\mathbf k}, i\omega_n)
\right) .
\end{equation}
This greatly speeded up the calculations
compared to adjusting $\mu$ at each iteration during the 
self-consistency cycle in order to maintain a given density
 $\langle n \rangle$.  
 Because the $\langle n \rangle$  changes
slightly at a function of $T$ for fixed $\mu$ this leads to small 
changes in density as a function of $T$.
For example the case where $t'=0$ and $\mu=.8$ the densities for
temperatures of $T=.03t$ and $T=.25t$ were $\langle n \rangle = 1.18$
and $\langle n \rangle = 1.19$ respectively.
These small changes in the density 
do not significantly alter our results.

In order to investigate the effects of the electronic band structure
of the cuprates we performed calculations for both
the nearest neighbor hopping only Hubbard model ($t'=0$),
and for a more realistic cuprate band structure of $t'=-0.3t$.
In the case $t'=0$ we concentrated our attention on the region near
to half filling, $\langle n \rangle =1$, and to an interaction strength 
of $U = 4t$. Exactly at half filling the 
Fermi surface is
perfectly nested along ${\mathbf Q}_{AF}$, and this also coincides with the
Van Hove singularity in the density of states. It is difficult therefore to
distinguish the relative importance of nesting and the Van Hove
singularity.
However, in the case where $t' \neq 0$  the Fermi surface does not have 
perfect nesting at any filling, and the Van Hove singularity occurs
at a finite doping. In this case we examined in detail both the region of
doping around half filling and the region around the Van Hove singularity.
We also compared the different effects of electron or hole  doping,
$\langle n \rangle>1$ or $\langle n \rangle<1$ respectively,
 corresponding to the n and p type
cuprate superconductors.

\section{ANALYSIS OF THE SPECTRAL FUNCTIONS}

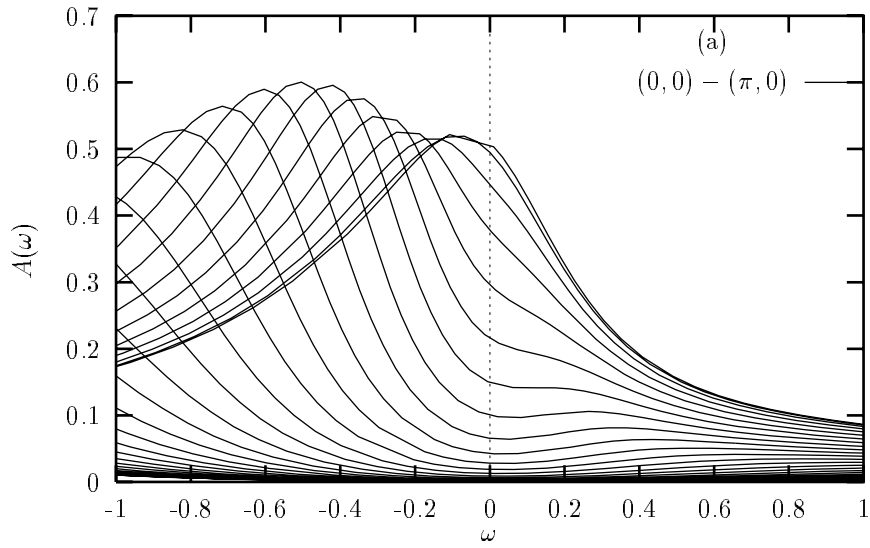
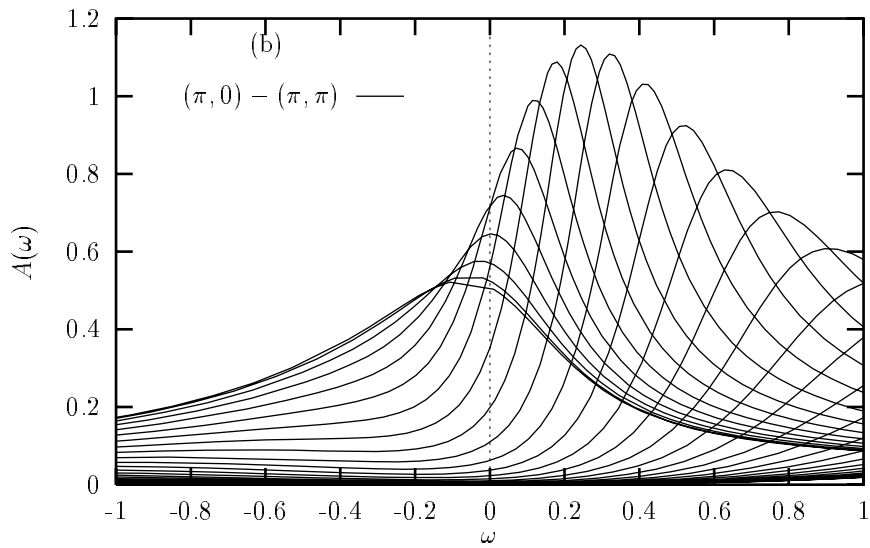
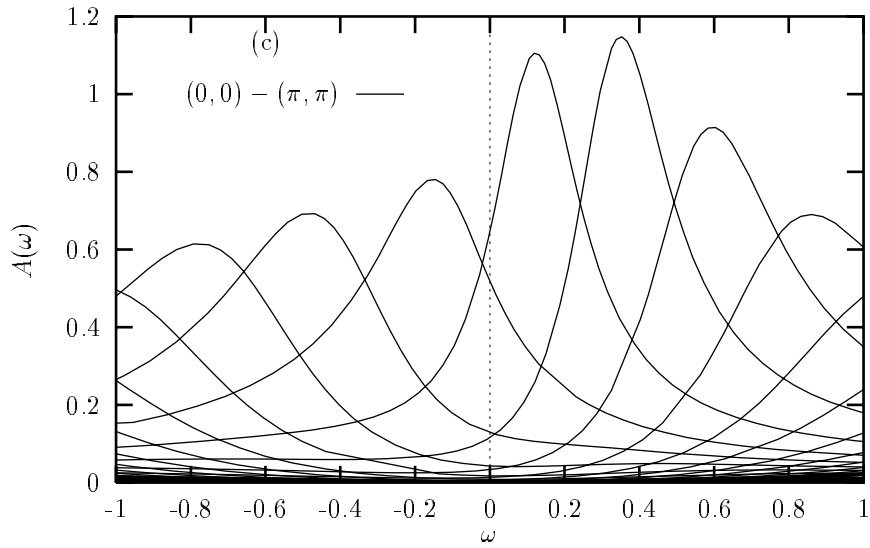
\begin{figure}
\caption{Spectral functions for momentum points along 
the lines $(0,0)-(\pi,0)$ (a) , $(\pi,0)-(\pi,\pi)$ (b) ,
and $(0,0)-(\pi,\pi)$ (c).
The calculations were for $t'= 0$ , $U=4t$ , $T = .1t$ and
$\langle n \rangle = 1.08$.}
\label{sfa}
\end{figure}

Figures \ref{sfa}(a,b,c)  show a typical series of spectral functions,
$A({\mathbf k},\omega)$. They 
illustrate the dispersion of the quasiparticle peak
along the Brillouin zone lines
${\mathbf k} = (0,0)- (\pi,0) $, ${\mathbf k} = (\pi,0)- (\pi,\pi) $, 
and ${\mathbf k} = (0,0)- (\pi,\pi) $.  In this case $t'=0$,
$U=4 t$, $T=.1 t$ and the band filling was
$\langle n \rangle=1.079$. The spectral functions in the figure show a 
rather broad quasiparticle peak which disperses as a function of
 ${\mathbf k}$. In Figs. \ref{sfa}(b,c) the peak 
narrows and then broadens again upon
passing through the Fermi surface. On the other hand, in  Fig. \ref{sfa}(a)
the peak is always broad. This is expected since
 the Fermi surface does not intersect
the line ${\mathbf k} = (0,0)- (\pi,0) $ at this band filling.
Overall the  qualitative behavior of the spectral function
is thus similar to that expected in a
Fermi liquid.

A  more precise quantitative analysis of the spectral functions
presents some difficulties.
While the quasiparticle
peak is quite well defined for most of the spectra, 
especially near the Fermi surface, in many cases the peak is very broad
and there is no clear separation between the coherent quasiparticle
peak and the incoherent background. The peaks tend to be asymmetric
rather than ideal Lorentzians.
It can also be seen in Fig. \ref{sfa}(b) that the sharpest peak
is not always centered at zero energy, the location of the Fermi surface.
 Furthermore one can also
see from the figure that that the peak centered at zero
energy does not have equal areas on either side of zero. This means that
this Fermi surface peak  does not correspond to the momentum where
the occupation number  equals one half.
The occupation number is essentially given by the
area under the graph at negative frequencies, 
\begin{equation}
 \langle n_{\mathbf k} \rangle = 
\int A({\mathbf k},\omega) f(\omega) d\omega \label{eq:areas}
\end{equation}
where $f(\omega)$ is the Fermi distribution.
Because of these ambiguities it is not even clear which peak
corresponds to the Fermi surface crossing.
The problem of locating the Fermi surface
from the spectra is discussed in the following section.
In the remainder of this section  we
concentrate on how to extract quantitative information
about the quasiparticle states from the
individual spectral functions.

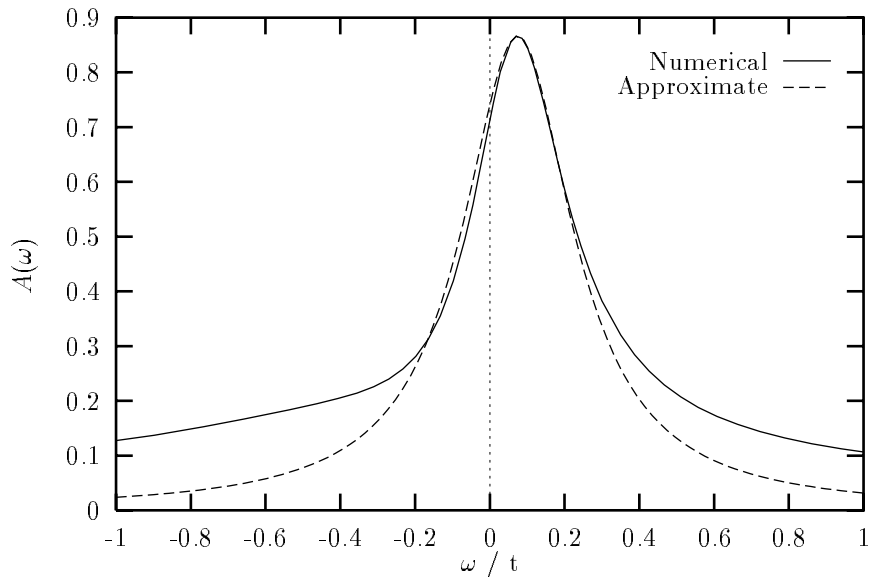
\begin{figure}
\caption{Spectral function $A({\mathbf k},\omega)$ for momentum point 
$(\pi , 5\pi / 32)$ from 
FLEX calculations compared  with that derived from the
 approximate form  of
Eq. \ref{approx}, where $\Gamma = .36737t$, $Z = .490$ and 
$\epsilon_{\mathbf k} = .074t$. These
values were taken from the height and the half width, using the
derivative of the Green's function, as in equation \ref{derivG},
 $Z = .477$.}
\label{sf}
\end{figure}

In order to make a more detailed analysis of the spectral functions
it is necessary to extract the  coherent quasiparticle
part of the spectrum defined by,
\begin{equation}
A({\mathbf k},\omega ) = {1\over \pi} \mbox{Im}
\left[ {1\over( \omega -
\epsilon_{\mathbf k})/Z_{\mathbf k} -i\Gamma_{\mathbf k}}
\right] .
\label{approx}
\end{equation}
Fitting the quasiparticle peak in the spectrum to this Lorentzian form
one can obtain
the quasiparticle band energy, $\epsilon_{\mathbf k}$, 
scattering energy, $\Gamma_{\mathbf k}$, and spectral
weight $Z_{\mathbf k}$.  A typical such Lorentzian fit
is shown in Fig. \ref{sf}.  Performing fits like this
on each spectral function $A({\mathbf k},\omega)$ defines the
parameters $\epsilon_{\mathbf k}$, $\Gamma_{\mathbf k}$ and
 $Z_{\mathbf k}$ throughout
the Brillouin zone. It is clear from Fig. \ref{sf} 
that the Lorentzian fit to the spectral function gives a good approximation
to the peak overall,  but the fitted Lorentzian decays faster than the
numerical function. There is also a slight asymmetry.
These differences are usually described as the
incoherent part of the spectrum.  
The peak is quite Lorentzian in shape and the  fitting parameters are 
generally well defined when 
the peak is narrow and symmetrical, but the fitting parameters
 are more ambiguous when the
spectral peak is broad or lop-sided.

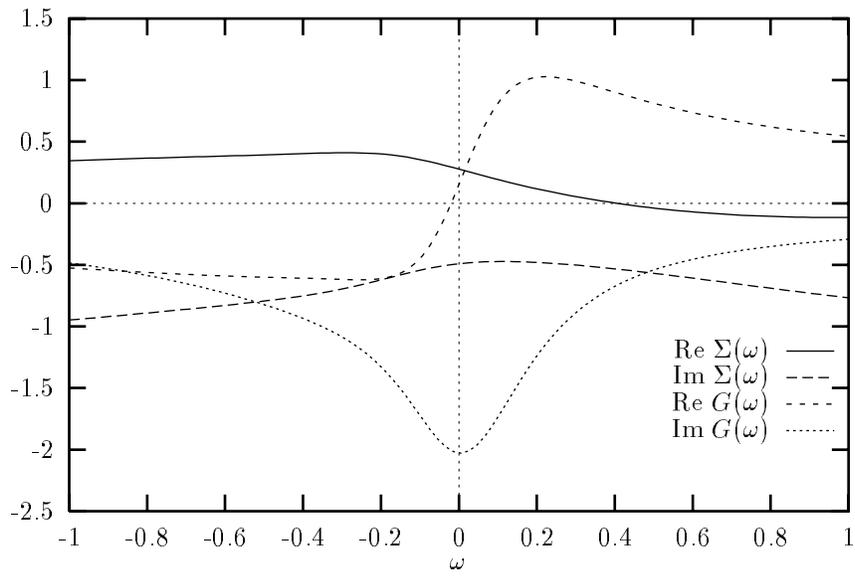
\begin{figure}
\caption{Plot of the real and imaginary components of the self-energy 
and the Green's function  for a momentum point of $(\pi , 3\pi / 32)$. 
This is the nearest momentum point to the Fermi surface crossing.
 The other parameters of the plot are 
$t' = 0$, $U = 4t$ , $T = .1t$ and $\langle n \rangle = 1.08$.
For this plot the real part of the Green's function is zero at a
frequency of $-.013t$, the minimum in the imaginary part of the Green's
function is at $.006t$ and the maximum in the imaginary part of the
self-energy is at $.106t$.
}
\label{sigma}
\end{figure}

In order to avoid these ambiguities in fitting the peak
we define the quasiparticle parameters in terms of the self-energy.
We calculate the real-axis self-energy from the
analytic continuation of the Green's function
\begin{equation}
\Sigma( {\mathbf k}, \omega )
 = \omega - \xi_{\mathbf k} - {1\over
G( {\mathbf k},\omega) } .
\end{equation}
This definition of $ \Sigma( {\mathbf k}, \omega )$ 
is more accurate than performing a separate
analytic continuation of the $\Sigma(i\omega_n)$ to the real axis.
Figure \ref{sigma} shows the Green's function and self energy for a 
point near the Fermi surface crossing.
As can be seen from the figure, the imaginary part of the
self-energy, although not constant, is slowly varying around the region
of the spectral function peak (where $-\mbox{Im}[G(\omega)]$ is maximum
in Fig. \ref{sigma}).
Also the real part of the self-energy is not precisely linear in
frequency over the width of the peak.  For these reasons the
spectral function peak is not an ideal Lorentzian.
In Fermi liquid theory one takes the limits
$\epsilon_{\mathbf k} \rightarrow 0$ and $T\rightarrow 0$
so that the quasiparticle peak is much narrower than the range of
frequencies over which $\mbox{Re}[\Sigma]$ and  $\mbox{Im}[\Sigma]$ vary.
However at the  temperatures and momentum resolutions we can access
 numerically
(and these correspond to reasonable physical temperatures in the cuprates,
eg. 100K) the quasiparticle peaks are wider than the energy scale
of variations in the self-energy. We shall show below that this leads to
significant deviations from Fermi liquid theory.

In order to avoid ambiguities arising from the Lorentzian fits 
we define the quasiparticle parameters
$\epsilon_{\mathbf k}$, $\Gamma_{\mathbf k}$ 
and $Z_{\mathbf k}$ from the self-energy.
The band energy is given by the solution of:
\begin{equation}
 \mbox{Re}[ \Sigma({\mathbf k},\epsilon_{\mathbf k})] =
 \epsilon_{\mathbf k} - \xi_{\mathbf k} , \label{eq:epsilon}
\end{equation}
which corresponds to the frequency where $\mbox{Re[}G({\mathbf k},\omega)]=0$
in Fig. \ref{sigma}.
The quasiparticle scattering rate is given by:
\begin{equation}
     \Gamma_{\mathbf k} = \mbox{Im}[\Sigma({\mathbf k},
\epsilon_{\mathbf k})],
\end{equation}
and  the quasiparticle spectral  weight is given by
\begin{equation}
Z_{\mathbf k}
 = {1 \over 1 -
 \mbox{Re}\left[ {\partial \Sigma \over \partial \omega } 
  \right] }.
\end{equation}
The latter can also
be written in terms of the Green's function as,
\begin{equation}
Z_{\mathbf k} = - \mbox{Re}\left[ { G(\omega)^2 \over \left(
 {\partial G(\omega )
 \over \partial \omega } \right) } 
 \right].
\label{derivG}
\end{equation}
Here the derivatives are evaluated at $\omega=\epsilon_{\mathbf k}$.
For the peak shown in Fig. \ref{sf} the fitted peak width led to a value
of $Z=0.490$ while the self-energy derivative led to a value of $Z=0.477$,
showing that the two definitions yield similar, but not identical, values.

\section{FERMI SURFACE LOCATION}

In order to discuss how $\Gamma_{\mathbf k}$ and 
$Z_{\mathbf k}$ vary on the Fermi surface we
need to have a method of precisely determining the position of the
Fermi surface. Unfortunately,
from the numerical data there are several definitions which
are possible, and which produce slightly different shapes.

Figure \ref{sigma} demonstrates that the maximum in the spectral
function, as shown by a minimum in the imaginary part of the Green's
function, does not coincide precisely with where the real part of the Green's
function is zero.  In other words the band energy, $\epsilon_{\mathbf k}$,
defined by the Lorentzian peak (Eq. \ref{approx}) differs slightly from
the one defined from the self-energy (Eq. \ref{eq:epsilon}). 
There is thus a small ambiguity in the location of the Fermi surface crossing.
In order to illustrate this point,
Fig. \ref{sigma} shows the momentum point $(\pi,3\pi/32)$. Here the
spectral function ($A\propto-\mbox{Im}[G]$) peak is centred at zero,
and so by Eq. \ref{approx} this momentum is a point in the Fermi surface.
On the other hand the real part of the Green's function is not quite
zero here, and so this point is not quite at the Fermi surface
according to Eq. \ref{eq:epsilon}.

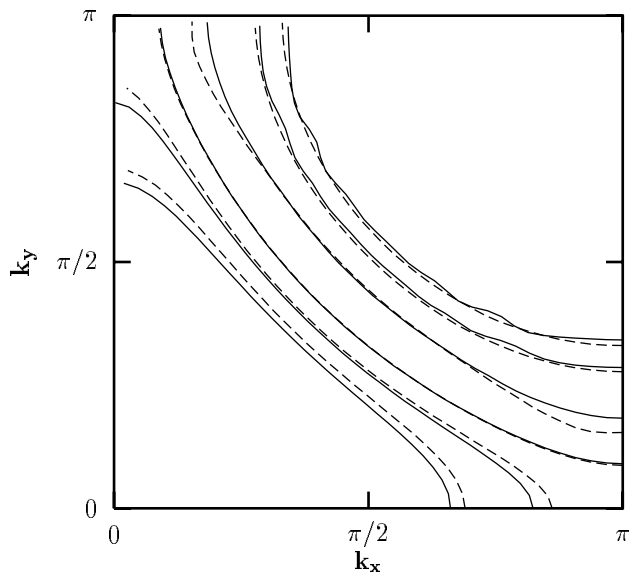
\begin{figure}
\caption{Fermi surfaces defined as $\langle n_{\mathbf k} \rangle = .5$,
solid lines, and $\epsilon_{\mathbf k} = 0$, dashed lines for densities
of $\langle n \rangle = .54, .61, .81, 1.01, 1.2, 1.3$.}
\label{td3fs}
\end{figure}

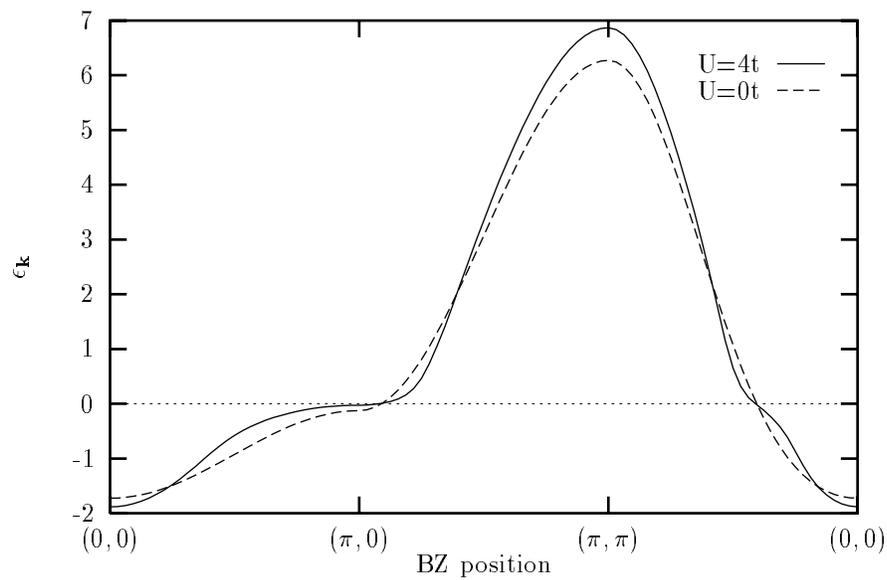
\begin{figure}
\caption{Band structure from $\epsilon_{\mathbf k}$, equation
\protect\ref{eq:epsilon} for $t' = -.3t$ , $T=.03t$ , $\langle n \rangle
 = .81$ and
$U = 4t$ and $U=0$.}
\label{bdtd3o12}
\end{figure}

A quite different definition of the Fermi surface location
can be made in terms of the momentum distribution function
$\langle n_{\mathbf k} \rangle$.   This can be evaluated directly from the
Matsubara frequency Green's function
\begin{equation}
\langle n_{\mathbf k} \rangle
 = {1\over 2} + \sum_{\omega_n} 
G(i\omega_n,{\mathbf k} ).
\end{equation}
and so (unlike Eq. \ref{eq:areas}) can be computed independently
of the Pad\'e analytic continuation to real frequencies.
In terms of $\langle n_{\mathbf k} \rangle$ 
the Fermi surface corresponds to the
discontinuity at zero temperature. However, at finite temperatures
$\langle n_{\mathbf k} \rangle$
 is continuous, and so it is difficult to precisely locate the 
 Fermi surface. A simpler definition, which 
can be evaluated much more reliably at finite temperatures is the choice
$\langle n_{\mathbf k} \rangle = 1/2$.
In practice we have found that this definition of
the Fermi surface is very similar to, but not identical with,
the Fermi surfaces defined by $\epsilon_{\mathbf k}=0$.
This can be seen in Fig. \ref{td3fs} where the Fermi surfaces arising
from the two different definitions, $\epsilon_{\mathbf k} = 0$ and
$\langle n_{\mathbf k} \rangle = 1/2$ are plotted for various densities.
The $\epsilon_{\mathbf k} =0$ Fermi surface is always shifted towards
the BZ diagonal $(\pi,0)-(0,\pi)$ as compared with the 
$\langle n_{\mathbf k} \rangle = 1/2$ Fermi surface, 
this is due to a flattening
of the quasi-particle band around the Fermi surface.
This flattening can be easily seen from the band structure as
derived from $\epsilon_{\mathbf k}$, equation \ref{eq:epsilon}.
Figure \ref{bdtd3o12} shows the band structure along the BZ path
$(0,0)-(\pi,0)-(\pi,\pi)-(0,0)$ for an interacting system, $U=4t$, and
that a non-interacting system with the same density.
The interacting band is flatter at the  Fermi energy as compared with
the non-interacting band. This is especially so  around the $(\pi,0)$
Van Hove saddle point.
The $(0,0)-(\pi,\pi)$ Fermi surface crossing also displays flattening of
the band, this makes the Fermi surface crossing a point of inflection in
the band structure.

In order to make a definite choice, in the remainder of this paper
we define the Fermi surface crossing to be the point where
$\epsilon_{\mathbf k}=0$,
 defining $\epsilon_{\mathbf k}$ from
the self-energy Eq. \ref{eq:epsilon}. We have found
that using other definitions leads to essentially the same qualitative 
behavior.
Our conclusions therefore do not depend on the specific definition.
In particular all of the results shown in the following section were obtained
for both the $\epsilon_{\mathbf k}=0$ and 
$\langle n_{\mathbf k} \rangle = 1/2$ definitions of the Fermi surface
and all the qualitative features found were very similar in both cases.

\section{TEMPERATURE DEPENDENT QUASIPARTICLE SCATTERING RATE}

We have calculated the Hubbard model spectral functions, quasiparticle
band parameters and Fermi surface at a wide range of 
temperatures and band fillings.  In this section we present our results
for the temperature dependence of the quasiparticle scattering rate
$\Gamma_{\mathbf k}$.  We especially focus on the scattering
rate on the Fermi surface, and its dependence on angle around
the Fermi surface, $\theta$. For a closed Fermi surface the angle $\theta$ 
was measured from $(0,0)$, and for an
open Fermi surface it was measured from $(\pi,\pi)$.

Below we consider in detail both the nearest neighbor only Hubbard
model $t'=0$,
and the next neighbor hopping model with $t'=-0.3t$.
As we shall show, the contrast between these two systems helps
clarify the different roles played by
Fermi surface nesting, antiferromagnetism and the Van Hove singularity.

\subsection{The $t'=0$ case}

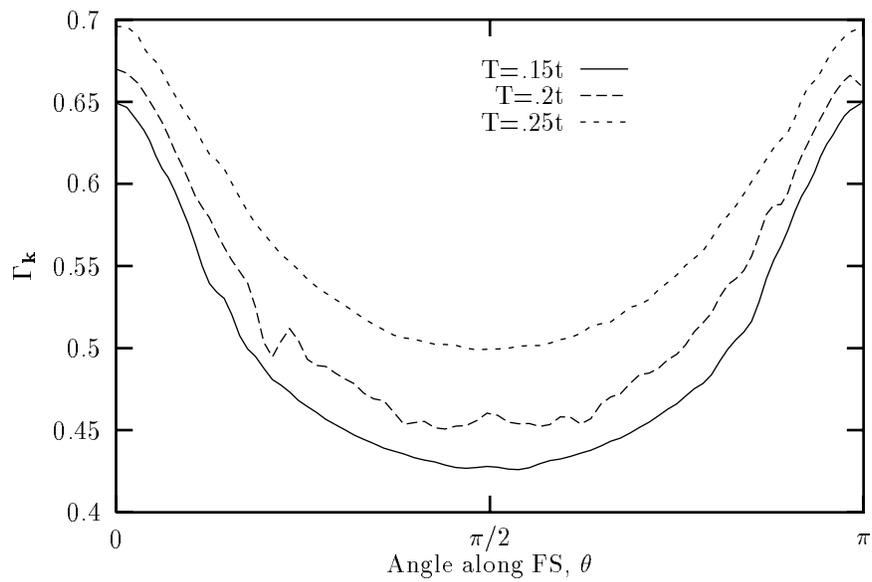
\begin{figure}
\caption{Scattering rate, 
$\Gamma_{\mathbf k}$, as a function of the angle along
the Fermi surface, $\theta$, 
for temperatures of $T=.15t$, $T=.2t$ and $T=.25t$.
Other parameters being $t'=0$, $U=4t$ and $\langle n \rangle =1 $.
}
\label{Gamma1}
\end{figure}

Fig. \ref{Gamma1} shows the quasiparticle scattering energy
at three different temperatures as a function of angle around the
Fermi surface. In this case the chemical potential was 
set to $\mu=0$ corresponding to half filling,  $\langle n \rangle=1$.
At this filling the Fermi surface is a perfectly nested square.
There are two interesting points from Fig. \ref{Gamma1}. Firstly 
$\Gamma_{\mathbf k}$ has a substantial anisotropy, being about
$50\%$ larger at angle $\theta=0$, the point $(\pi,0)$, than
at $\theta=\pi/2$, the point $(\pi/2,\pi/2)$. This implies that
the states near the Van Hove singularity are more heavily scattered
than the states near the center of the nested pieces of Fermi surface.
The second observation one can make from Fig. \ref{Gamma1} is that
the scattering rate decreases rather weakly with temperature. On lowering the
temperature from $T=0.25t$ to $T=0.15t$ $\Gamma_{\mathbf k}$ decreases
 by only
about $10-20\%$. This is clearly a non-Fermi liquid temperature dependence.
We cannot explore this effect at lower temperatures for this band
filling, since the system becomes antiferromagnetic.

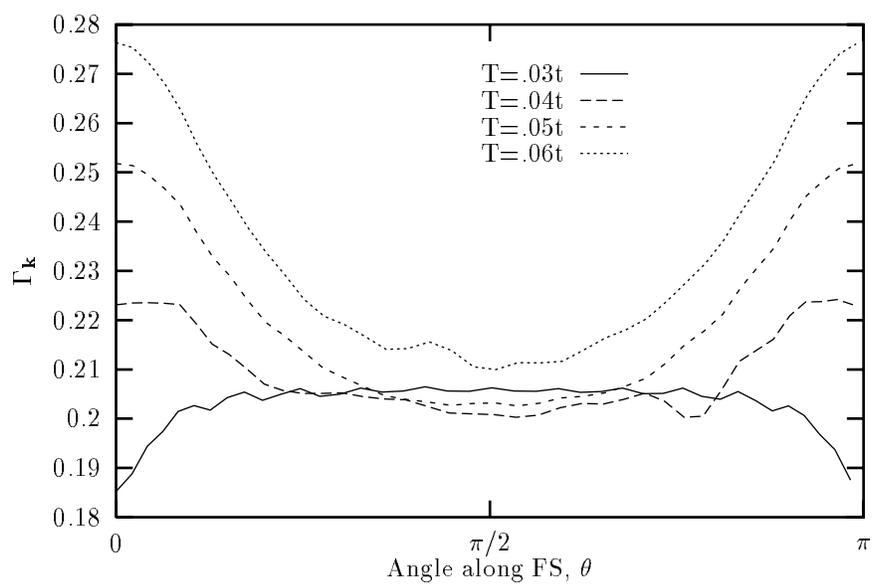
\begin{figure}
\caption{Scattering rate, $\Gamma_{\mathbf k}$, as a 
function of the angle along
the Fermi surface, $\theta$, for temperatures of $T=.03t$ to $T=.06t$.
Other parameters being $t'=0$, $U=4t$ and $\mu = .6t$ 
($\langle n \rangle \approx 1.13$) .
In this plot the scattering energy for $T=0.3t$ is higher around the
$\pi/2$ angle that that for higher temperatures, this is typical
`hot spot' behavior.
}
\label{Gamma2}
\end{figure}

In order to go to lower temperatures in the paramagnetic phase it
is necessary to go off half filling. Fig. \ref{Gamma2} 
shows the temperature dependent $\Gamma_{\mathbf k}$
on the Fermi surface corresponding to $\mu=0.6t$
( $\langle n \rangle \approx 1.13$). Now both the angular dependence
and temperature dependence are quite different.  At high temperatures
the scattering is maximum around $\theta=0$, similar to Fig. \ref{Gamma1}.
However at lower temperatures the maximum scattering occurs at
$\theta=\pi/2$. In fact at $\theta=0$ the temperature dependence of
$\Gamma_{\mathbf k}$
is almost linear.  On the other hand, at $\theta=\pi/2$ $\Gamma_{\mathbf k}$
 is only
weakly temperature dependent. In fact $\Gamma_{\mathbf k}$ first decreases
with decreasing $T$, but then it has a minimum and starts to increase
again.  We shall refer to this anomalous  behavior as `hot spot'
behavior, for reasons which will be clear below. Notice that
this behavior allows us to define a temperature scale, $T^*$,
from the position of the minimum in $\Gamma_{\mathbf k}$ as a function of $T$.
For example $T^*\approx 0.45t$ in Fig. \ref{Gamma2}.
This $T^*$ may be related to the physical pseudogap behavior seen in
ARPES and other experiments in the cuprates. However we shall
always define our $T^*$ from the temperature of the 
minimum in $\Gamma_{\mathbf k}$
rather than from any
other specific features in the spectral functions.

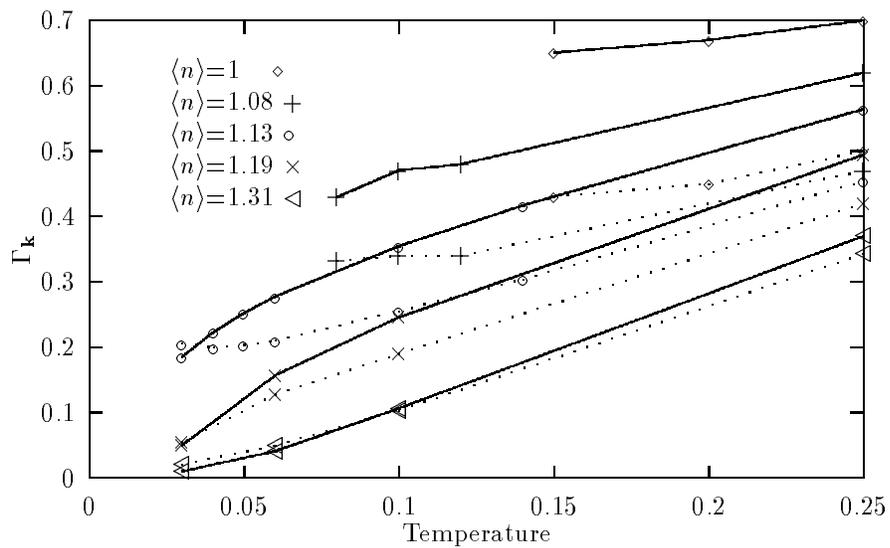
\begin{figure}
\caption{Temperature dependence of the scattering rate, 
$\Gamma_{\mathbf k}$, for
$t'=0$, $U=4t$, and an angle along the Fermi surface of $\theta = 0$ 
and $\theta = \pi/2$.
The plots for various densities show that the temperature dependence is
nearly linear for a wide range of densities with a positive intercept at
zero temperature for densities below $\langle n \rangle = 1.19$.
}
\label{TGamma3}
\end{figure}

Fig. \ref{TGamma3} summarizes the temperature dependence of
$\Gamma_{\mathbf k}$ at the Fermi surface angles $\theta=0$
and $\theta=\pi/2$ and at different band fillings.
One can clearly see that that the upturn in $\Gamma_{\mathbf k}(T)$ 
for $\theta=\pi/2$ occurs at a range of band fillings, but that
$T^*$ decreases on moving away from half filling. The upturn appears
to occur just before entering the antiferromagnetic phase
so that $T^* > T_{AF}$.  Beyond about
$\langle n \rangle>1.15$ there is no antiferromagnetism and
no upturn at all.  

Well away from half filling, at $\langle n \rangle=1.31$, the scattering
rate $\Gamma_{\mathbf k}$ becomes isotropic. 
Here the temperature dependence is  linear down to a 
 tempertaure of about $T=0.06t$.  Assuming that $t\approx 0.5$eV in the
 cuprates this would
correspond to a temperature of 300K. 
Below this temperature there is some evidence
 of a cross over to a $T^2$ behavior, 
suggesting that the system is indeed a Fermi liquid. 
The
characteristic temperatures for the observation of Fermi liquid behavior
are  very much lower than either $t$, or $J = 4t^2/U$.  
A linear temperature dependence of $\Gamma_{\mathbf k}$
 occurs at a wide range of
dopings for the Fermi surface angle $\theta=0$. Since this is the
point where the Fermi surface is closest to the Van Hove saddle point, 
Fig \ref{TGamma3} suggests
that the linear $T$ dependence of $\Gamma_{\mathbf k}$ may be associated
 with the proximity of the Fermi energy to a Van Hove singularity.

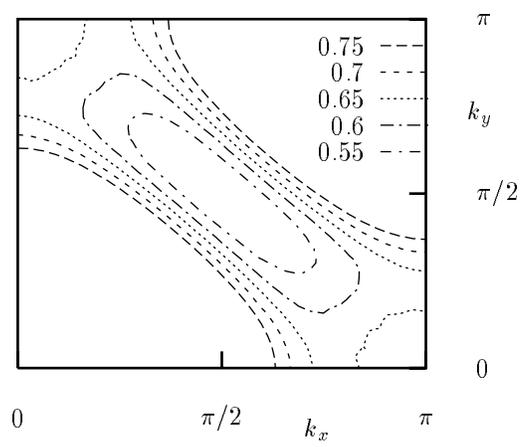
\begin{figure}
\caption{Contour plot of the scattering energy, $\Gamma_{\mathbf k}$, 
as a function
of Brillouin zone position, for $t'=0$, $U=4t$, $T=.25t$ and $\langle n
\rangle= 1$.
The saddle points at $(\pi,0)$ and $(0,\pi)$ are not at a minimum of the
scattering energy and show enhanced scattering there.
In this plot the contours are only shown in the region close to the
Fermi surface, which in this case is 
the diagonal from $(\pi,0)$ to $(0,\pi)$.
}
\label{Gamma3d1}
\end{figure}

In order to gain further insight into the anisotropy in the 
scattering rate,
Fig. \ref{Gamma3d1}  shows a contour plot of  $\Gamma_{\mathbf k}$, 
for a quadrant of the BZ at half filling.
Here it is seen that the scattering rate actually has a maximum on
the FS at the point ($\pi,0$). This is definitely unlike a Fermi liquid,
for which $\Gamma_{\mathbf k}$ goes through a minimum on crossing
the Fermi surface.  One can see from Fig. \ref{Gamma3d1} that
this Fermi liquid  minimum of the scattering rate is
seen for crossing the FS at the $(\pi/2,\pi/2)$ point.

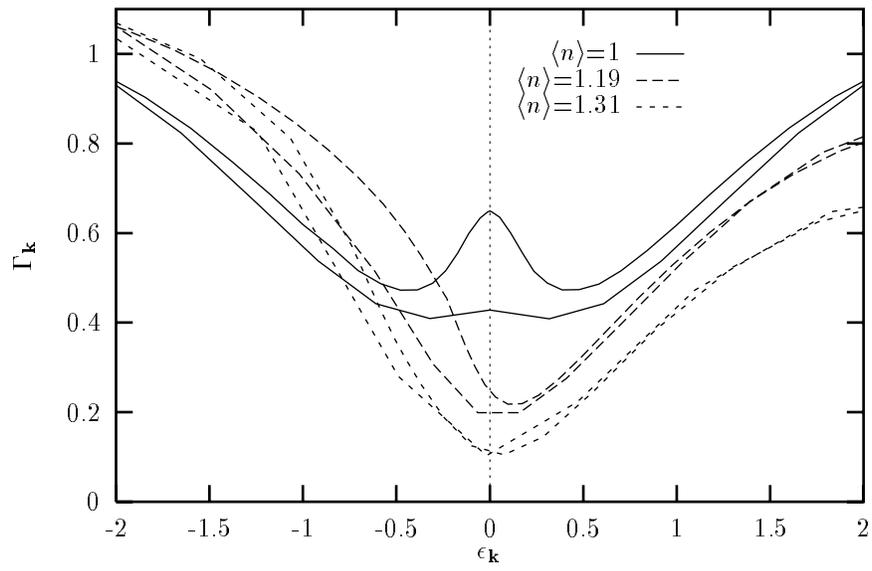
\begin{figure}
\caption{Plot of the scattering rate, $\Gamma_{\mathbf k}$,
versus  the  band energy, $\epsilon_{\mathbf k}$, 
 for 
momentum points along the path $(0,0)-(\pi,0)-(\pi,\pi)-(0,0)$ and with
$t'=0$, $U=4t$,
$T=.1t$ and densities of $\langle n \rangle = 1, 1.19, 1.31$.
The two lines for each plot roughly correspond to the maximum and
minimum values of the scattering rate for a particular energy.
These plots show clear deviation from Fermi liquid behavior since the
scattering rate is not proportional to the square of the band energy.
The separation of the lines gives an indication of the anisotropy 
of the scattering rate.
For the $\langle n \rangle = 1.31$ plot the scattering rate  no
longer has a large anisotropic region and although the energy 
dependence is steeper on
the negative energy side the behavior is more Fermi liquid like.
}
\label{BandGamma1}
\end{figure}

For a Fermi liquid one also expects the scattering rate to vary as the square
of the band energy.
Fig. \ref{BandGamma1} shows the scattering rate as a function
of the band energy for densities of $\langle n
\rangle=1$,
$\langle n \rangle=1.19$ and $\langle n \rangle = 1.31$.
For each density we have plotted both the maximum and minimum scattering
rates $\Gamma_{\mathbf k}$ for a given energy $\epsilon_{\mathbf k}$.
The anisotropy of the scattering energy is then given by the 
difference between these two curves.
The figure shows that the anisotropy is centered around zero energy 
for the half-filling
 case, but moves to lower energy as the density is increased.
This  indicates that
  the anisotropy is associated with the Fermi surface
proximity to the BZ diagonal or the Van Hove singularity.
Once the FS has been moved away from this anisotropic region the energy
dependence becomes $\omega^2$ like.
From these plots we determine that the system resembles the Fermi liquid
model for densities above and including $\langle n \rangle=1.31$.

\subsection{The $t'=-.3t$ case}

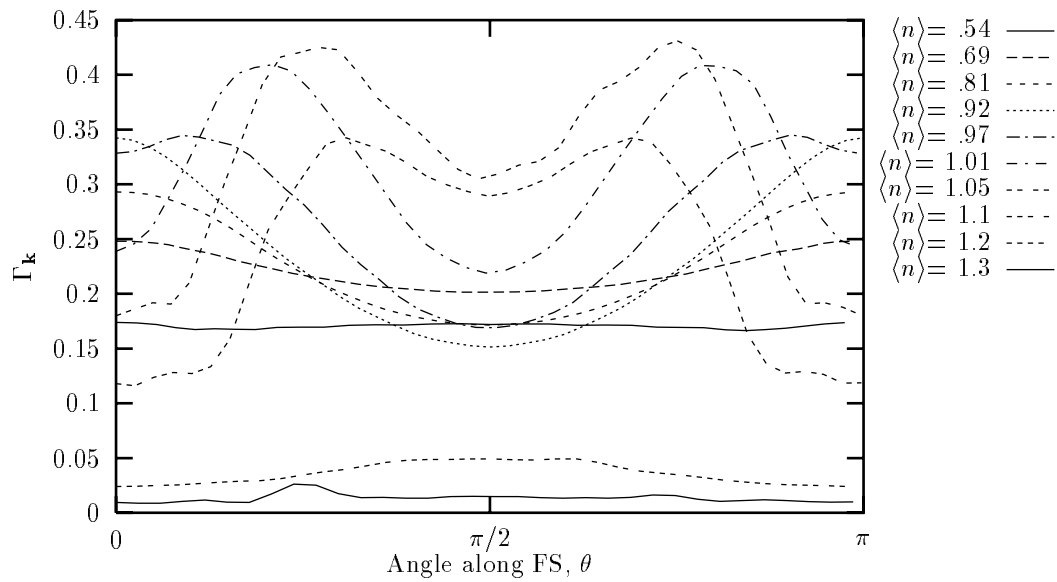
\begin{figure}
\caption{Scattering rate, $\Gamma_{\mathbf k}$, 
as a function of the angle along
the Fermi surface, $\theta$, for densities from $\langle n \rangle = .54$ 
to $\langle n \rangle = 1.3$, other
parameters being $t'=-.3t$, $U=4t$, and $T=.06t$.
The scattering rate is maximum for $\langle n \rangle = 1.05$ while the
antiferromagnetic ordering region occurs between $\langle n \rangle = 1.1$
and $\langle n \rangle
= 1.2$.
}
\label{Gamma3}
\end{figure}

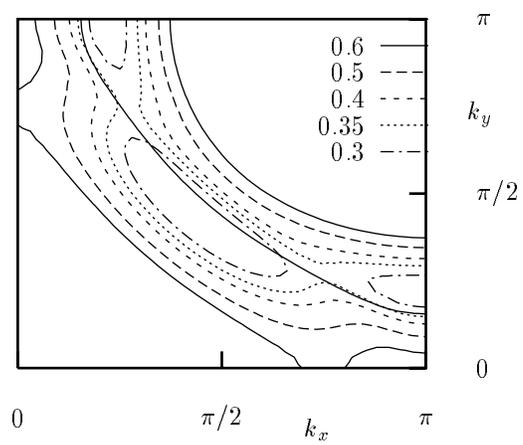
\begin{figure}
\caption{Contour plot of the scattering rate, $\Gamma_{\mathbf k}$, 
as a function
of Brillouin zone position, for $t'=-.3t$, $U=4t$, $T=.1t$ and $\langle n
\rangle= 1$.
The `hot spots' can be clearly seen on the diagonal of the BZ.
The solid crossing line is the Fermi surface.
}
\label{Gamma3d2}
\end{figure}

We will now compare the results with those from the case when
 $t' = -.3$.
For this case the point where the Fermi surface is at the band saddle point
is now at a density of $\langle n \rangle =.75$.
Figure \ref{Gamma3} shows the scattering rate, $\Gamma_{\mathbf k}$,  
along the Fermi surface
for various densities at a temperature of $.06t$. It can be seen that 
the scattering rate has a maximum between an angle of $\theta = 0$ 
and $\theta = \pi/2$  unlike
the $t'=0$ case. 
Figure \ref{Gamma3d2} shows a contour plot of the scattering rate,
$\Gamma_{\mathbf k}$, for a density of  
$\langle n \rangle
=1.01$ and a temperature of $T=.03t$.
The maximum in the scattering rate is seen to correspond to the point
 where the FS
crosses the BZ diagonal, as previously noted by Altmann Brenig and
Kampf\cite{kampf}.

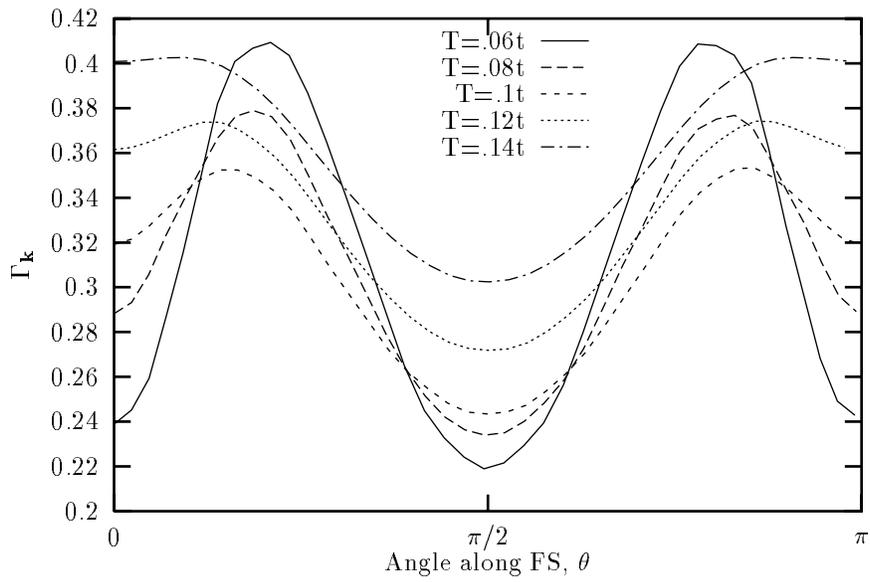
\begin{figure}
\caption{Scattering rate, $\Gamma_{\mathbf k}$, as a function of the 
angle along
the Fermi surface, $\theta$, for temperatures from $T=.06t$ to $T=.14t$, other
parameters being $t'=-.3t$, $U=4t$, $\mu = -.4$ ( $\langle n \rangle
\approx 1.01$ ).
The `hot spots' can clearly be seen at  angles of $\pi /4$ and
$3\pi / 4$, where the scattering rate increases with lowering
temperature.
}
\label{Gamma5}
\end{figure}

Looking at the temperature dependence of this maximum we can see, Fig.
\ref{Gamma5}, that  $\Gamma_{\mathbf k}$ increases with decreasing
 temperature. I.e.  it can be described as a
`hot spot', and shows a pseudogap $T^*$ behavior.
 For the lowest temperature, $T=0.06t$, the maximum in the
scattering rate, at angles of about $\theta = \pi/4$ and $\theta = 3\pi/4$, 
is almost
double the value at the minima, at angles of $\theta = 0$ and 
$\theta = \pi/2$.
This enhancement in the scattering rate can also be called a
`hot spot' in the sense  that it
is confined to a small region of the Fermi surface, 
consistent with the model of Pines and co-workers\cite{pines}.

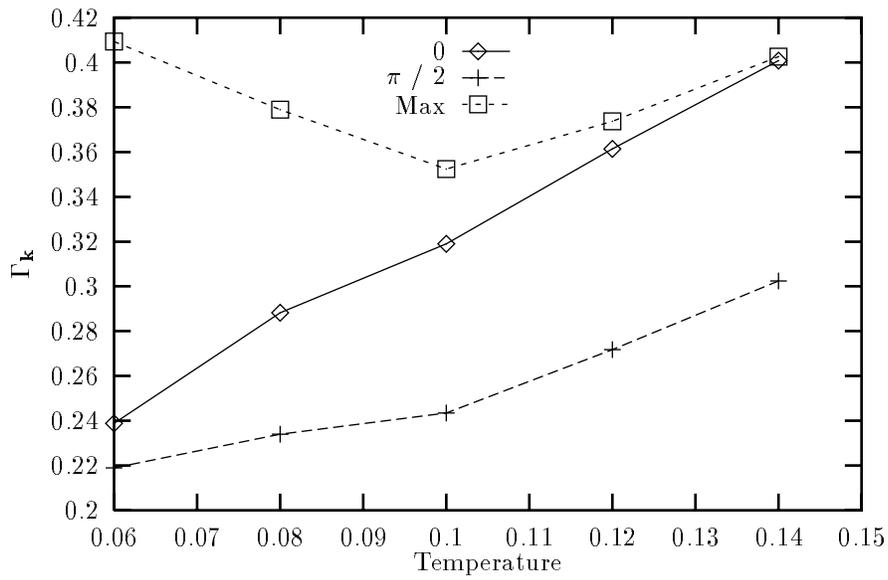
\begin{figure}
\caption{Temperature dependence of the scattering rate, 
$\Gamma_{\mathbf k}$, for
$t'= -.3t$ , $U=4t$ , $\langle n \rangle = 1.01$. The plots drawn show the
scattering rate at angles along the Fermi surface
 of $\theta = 0$ and of $\theta =\pi/2$ and also shown is the maximum
scattering rate, which for this case occurs at an angle of about
$\theta = \pi / 4$. At this angle the scattering rate 
increases with decreasing
temperature, evidence for a 'hot spot' on the Fermi surface.
}
\label{TGamma1}
\end{figure}

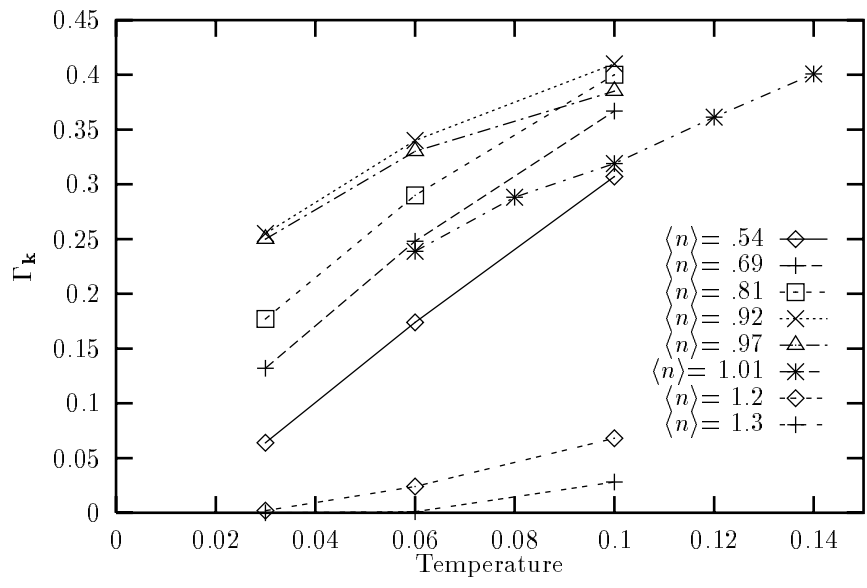
\begin{figure}
\caption{Temperature dependence of the scattering rate, 
$\Gamma_{\mathbf k}$, for
$t'=-.3t$, $U=4t$, and an angle along the Fermi surface of $\theta =0$.
The plots for various densities show that the temperature dependence is
nearly linear for a wide range of densities with a positive intercept at
zero temperature for all densities between 
$\langle n \rangle = .54$ and $\langle n \rangle = 1.2$.
}
\label{TGamma2}
\end{figure}

Figure~\ref{TGamma1} shows the temperature dependence of the
scattering rate at angles of $\theta = 0$ and $\theta = \pi/2$ and 
for the maximum value.
The scattering rate at $\theta=0$ and $\theta=\pi/2$
 continue to decrease roughly 
linearly with temperature, and behave quite differently from the $T^*$
upturn seen at the `hot spots'.
It also apears that at an angle of $\theta = 0$ and $\theta  = \pi/2$ the 
scattering rate will have a positive intercept
at zero temperature, although the zero temperature values cannot be
obtained since the system becomes an AFM below $T=.04t$.
Plotting the temperature dependence of $\Gamma_{\mathbf k}$ at
 $\theta=0$  for various
densities, see figure~\ref{TGamma2}, it is seen that for densities
between $\langle n \rangle =.69$ and $\langle n \rangle=1.01$ inclusive 
the
temperature dependence is always roughly
linear or sublinear with a positive intercept at zero temperature.
The positive intercept indicates that the scattering energy is unlikely
 to cross over to a Fermi liquid $T^2$
dependence at low temperatures.
This might be due to an AFM phase at lower temperatures which we cannot
 access, or it might indicate a non-Fermi liquid ground state.
However for
densities outside the region $ 0.69 < \langle n \rangle < 1.01$
we can say that although the high
temperature dependence of the scattering energy is linear, the low
temperature dependence could be $T^2$. In fact for the densities above
$\langle n \rangle=1.2$ this $T^2$ behavior can already be seen
quite clearly in Fig.~\ref{TGamma2}.

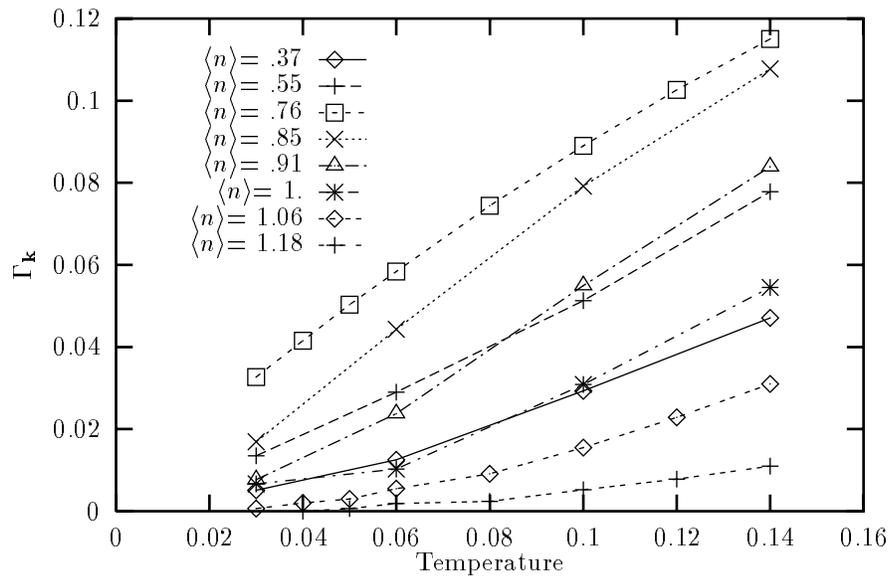
\begin{figure}
\caption{
Temperature dependence of the scattering rate, $\Gamma_{\mathbf k}$, for
$t'=-.3t$, $U=2t$, and an angle along the Fermi surface of $\theta =0$.
The plots for various densities show that the temperature dependence is
nearly linear only for a few values of the density with the maximum 
scattering occurring at $\langle n \rangle = .76$. 
At this density the Fermi energy is at the Van Hove singularity in the
density of states.
}
\label{gamtd3u2}
\end{figure}

All of the results described so far were obtained 
using the interaction, $U=4t$. For this value of $U$
it is difficult to separate the origins of the  
linear temperature
dependence of $\Gamma_{\mathbf k}$ near $\theta=0$
from  the effects of the `hot spots' and the proximity to the
AFM transition, since they occur at similar doping levels. 
However, lowering the interaction to $U=2t$ we are able to 
separate the various competing physical effects. We can then see
that the linear temperature dependence is associated with the Van Hove
crossing point and not AFM.
Figure \ref{gamtd3u2} shows the temperature dependence  of the
 scattering energy for various densities with $U=2t$.
The density region where there is a linear temperature
dependence has now been reduced to a small region around a density of
$\langle n \rangle = .76$ which corresponds to the point where the
Fermi energy is on the Van Hove singularity in the density of states.
For densities above $\langle n \rangle = .91$ and below $\langle n 
\rangle = .55$ there is a $T^2$ behavior, 
or a  least a positive curvature at low temperature.
For half filling the temperature dependence shows a slight up turn at
low temperatures, indicating a `hot spot' with a small $T^*$,
although no AFM phase is observed down to a
temperature of $T=.03t$. There are no hot spots at any other filling
for $U=2t$.

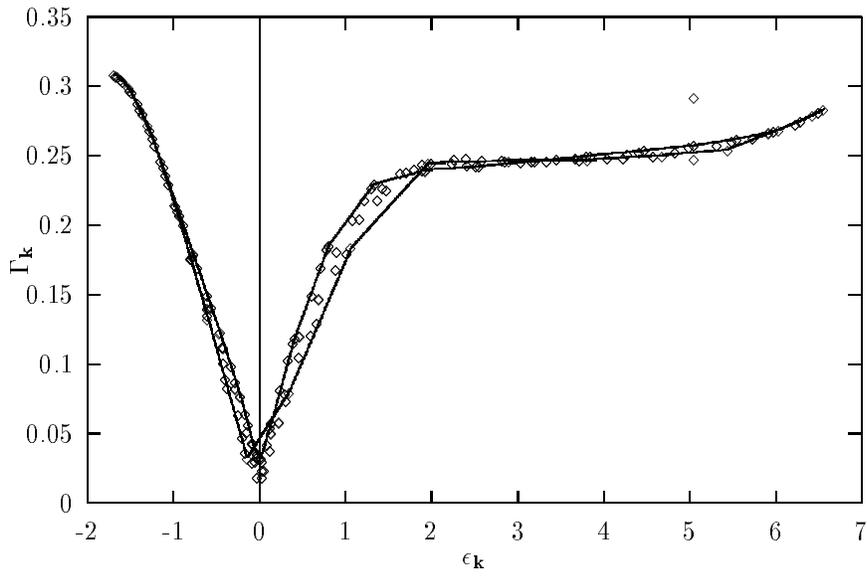
\begin{figure}
\caption{Plot of the scattering rate, $\Gamma_{\mathbf k}$, as a
function of band energy, $\epsilon_{\mathbf k}$, for all
momentum points (dots)  for $t'=-.3$, $U=2t$,
 $\langle n \rangle = .76$ and $T=.03t$.
The line connects the points on the path $(0,0)-(\pi,0)-(\pi,\pi)-(0,0)$
in the BZ.}
\label{BGamma5}
\end{figure}

As well as the scattering rate displaying linear temperature dependence
for densities around $\langle n \rangle = .76$ the dependence of the
scattering rate with band energy, $\epsilon_{\mathbf k}$, is also
linear.
This can be seen in Figure \ref{BGamma5} which is for $\langle n \rangle
= .76$ and $U=2t$. For $U=4t$  the energy dependence of the
scattering rate is masked by the 
increased anisotropy of the system.
Despite this the energy dependence of $\Gamma_{\mathbf k}$ 
apears to be linear when the
temperature dependence is also linear.

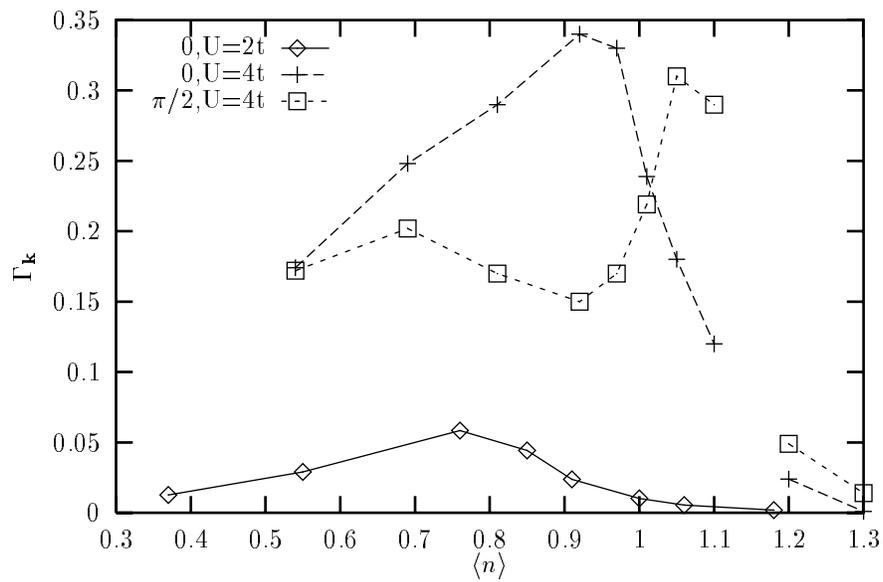
\begin{figure}
\caption{
Scattering rate, $\Gamma_{\mathbf k}$, 
as a function of density for an interaction
strength of $U=4t$ at an angle along the Fermi surface of $\theta =0$ and
$\theta = \pi/2$ and for an interaction strength of $U=2t$ at an angle
 along the
Fermi surface of $\theta =0$. Other parameters being $t'=-.3t$ and $T=0.6t$.
}
\label{td3t06gm}
\end{figure}

The influence of the Van Hove singularity can be most clearly 
seen in a plot of the
scattering rate as a function of doping, Fig. \ref{td3t06gm}.
In this plot the maximum in the scattering energy for $U=2t$, which 
occurs at
$\langle n \rangle = .76$, is also
mirrored by a peak in the scattering energy for $U=4t$.
Although the $U=4t$ plots are
dominated by the enhancements from the Fermi surface `hot spots'
occuring near half filling.

\section{Conclusions}

In our calculated quasiparticle
scattering rate $\Gamma_{\mathbf k}$ we observe two distinct 
non-Fermi liquid phenomena:
regions of linear in temperature dependence of $\Gamma_{\mathbf k}$,
and  'hot spots', a localized increase in the scattering rate with
decreasing temperature.  Fig. \ref{phase} shows the regions of 
density and temperature at which these two separate phenomena can
be observed.  

The top panel to Fig. \ref{phase} shows relevant regions
in the  phase diagram of the 
nearest neighbor Hubbard model with $t'=0$ and $U=4t$.  
The system is symmetrical around half filling and
the AFM phase extends to $\langle n \rangle \approx 
1 \pm .13$.  Near the boundaries of the antiferromagnetic
region we observe `hot spot' behavior, and the corresponding
pseudogap temperature $T^*$ of the upturn in $\Gamma_{\mathbf k}$ 
is marked on the figure.
However one can see from Fig.~\ref{phase} that Fermi liquid behavior
is not recovered ouside the region of antiferromagnetism
until a density of more than $0.3$ away from half filling. 
For densities between the AFM phase and 
$\langle n \rangle 1 = \pm 0.3 $ $\Gamma_{\mathbf k}$ is dominated
by a linear temperature dependence. 

The second panel in Fig.~\ref{phase} shows how this behavior is modified
by a non-zero $t'$ ($t'=-0.3t$) and the same value of $U$.
In this case the phase diagram is no longer symmetrical. The 
AFM region is centred just above half-filling, so that the AFM 
phase is destroyed much more quickly with p-type as opposed to n-type
doping, similar to the experimental behavior in the cuprates.
The `hot spot' behavior is now confined to the p-type side of the
AFM phase, and Fermi liquid behavior begins immediately
after the antiferromagnetism is destroyed for n-type doping.
Below half filling there is a wide region of marginal Fermi liquid
behavior (MFL on the figure) corresponding to a linear
temperature dependence of $\Gamma_{\mathbf k}$. This MFL 
region is centered around the Van Hove singularity and extends down 
as far as $\langle n \rangle \sim 0.5$.

Finally, the bottom panel in Fig.~\ref{phase} shows the 
behavior at the smaller value of $U=2t$ and the same value of
$t'=-0.3t$. Here the antiferromagnetism has essentially
disappeared, as has the `hot spot' behavior. On the other hand there
is still a MFL region, which is very clearly centered on the 
van Hove singularity at $\langle n \rangle = 0.75$.

Our calculations have ramifications for a number of the current theories
of high temperature superconductivity. Firstly the observations
of `hot spots', with a corresponding pseudogap temperature scale, $T^*$,
supports the concepts of {\it hot-} and {\it cold-quasiparticles} 
as advocated by Pines and others\cite{pines}. We find that
this behavior is a consequence of the proximity to the AFM phase,
and would be especially pronounced in the underdoped p-type cuprates
(assuming that the physical parameters are roughly similar to
those of the central panel in Fig.~\ref{phase}), as is indeed the case
experimentally. 
From the calculations described in this paper we indeed see
that the `hot spots' are associated with
quasiparticles with increased scattering located near the singular points
on the Fermi surface connected by the AFM wave vector $Q_{AF}$,
as also noted  previously by Altmann, Brenig and Kampf\cite{kampf}.

On the other hand, the `hot spot' behavior appears to be unrelated to
the  MFL behavior ($\Gamma_{\mathbf k} \sim T, \epsilon_{\mathbf k}$)
which we also observe. 
The parameter region where these {\it hot-quasiparticles} occur 
is confined
to a small region near to the AFM transition, and
the temperature dependence of the scattering rate of the `hot spot'
quasiparticles is not  linear in temperature.
We observe MFL
behavior over a much wider region of parameter space than the `hot spots'.
For $U=4t$ the MFL and hot spot behavior may
occur at the same band fillings, but on different pieces of
Fermi surface. However  for band fillings further away from the
AFM phase only MFL is observed.
For $U=2t$ both the AFM and the `hot spot' behavior
are suppressed, showing clearly that the MFL behavior
originates in the proximity to the Van Hove singularity
not the AFM.
This observation is in agreement with the scattering rate
predictions of
the 'Van Hove scenario' theories of the cuprates\cite{newns2}.

Note that we have only performed normal-state
calculations, and so have not included any possible d-wave 
superconducting phases in Fig.~\ref{phase}. We cannot therefore comment on
the existence of superconductivity induced by spin fluctuations near
 to the AFM phase
boundary\cite{pines} or whether the Van Hove singularity
also plays a role in enhancing the superconductivity\cite{newns2}.

In summary we find that the Van Hove singularity produces marginal
Fermi liquid behavior, scattering rates linearly dependent on
temperature,
 while the AFM phase is associated with {\it
hot-quasiparticles}. These {\it hot-quasiparticles} are characterised by
increasing scattering rates with decreasing temperature,
have a characteristic pseudogap temperature scale $T^*$, and are
localized to a small region of the BZ near the points where the
Fermi surface crosses the diagonal and are connected by the antiferromagnetic
 wave vector
$Q_{AF}$.

\acknowledgments

We would like to thank B.L. Gy\"{o}rffy for stimulating discussions.
This work has been supported by the Office of Naval Research Grant No.
N00014-95-1-0398.

\begin{figure}
\caption{
Rough sketch of the phase diagram, showing the regions of temperature
and density at which we see Fermi liquid behaviour (FL), 
marginal Fermi liquid behavior (MFL), AFM ordering
and $T^*$ the characteristic temperature at which `hot spots' occur.
The top graph is for $t'=0$ and $U=4t$, the middle graph is for
$t'=-.3t$ and $U=4t$ and the bottom graph is for $t'=-.3t$ and $U=2t$.
}
\label{phase}
\end{figure}

\begin{references}
\bibitem{hubbard} J. Hubbard, Proc. Roy. Soc. (London) {\bf A276},
238 (1963).
\bibitem{brinkman} W.F. Brinkman and T.M. Rice, Phys. Rev. B {\bf 2}, 
4302 (1970).
\bibitem{hirsch} J.E. Hirsch, Phys. Rev. B {\bf 31}, 4403 (1985).
\bibitem{gross} C. Gross, R. Joynt and T.M. Rice, Z. Phys. B {\bf 68},
425 (1987).
\bibitem{baskaran} G. Baskaran, Z. Zou and 
P.W. Anderson, Solid State Commun. {\bf 63}, 973 (1987).
\bibitem{anderson} P.W. Anderson, Phys. Rev. Lett. {\bf 64}, 1839 (1990).
\bibitem{affleck} I. Affleck and B. Marston, Phys. Rev. B {\bf 37},
3774 (1988).
\bibitem{varma} C.M. Varma {\it et al.}, Phys. Rev. Lett.
{\bf 63}, 1996 (1989).
\bibitem{annett} J.F. Annett, N. Goldenfeld and A.J. Leggett,
in ``Physical Properties of High Temperature Superconductors V'',
D.M. Ginsberg (ed.) (World Scientific, Singapore 1996).
\bibitem{bacci} S.B. Bacci, E.R. Gagliano, R.M. Martin and
J.F. Annett, Phys. Rev. B {\bf 44}, 7504 (1991).
\bibitem{bickers}
N.E Bickers and D.J. Scalapino, Ann. of Phys. {\bf 193}, 206-251 (1989).
\bibitem{BSW}
N.E.Bickers, D.J.Scalapino and S.R.White, Phys. Rev. Lett. {\bf 62},961
(1989).
\bibitem{DHS}
J.J.Deisz, D.W.Hess and J.W.Serene, Phys.Rev.Lett. {\bf 76},1313 (1996).
\bibitem{BW}
N.E.Bickers and S.R.White, Phys.Rev.B {\bf 43}, 8044 (1991).
\bibitem{Yone}
Kenji Yonemitsu, J. Phys.Soc.Jpn. {\bf 58} 4576 (1989).
\bibitem{Baym}
Gordon Baym, Phys.Rev. {\bf 127}, 1391 (1962).
\bibitem{pines} A. Sokol and D. Pines, Phys. Rev. Lett. {\bf 71},
2813, (1993); P. Monthoux and D. Pines, Phys. Rev. B {\bf 50},
16015 (1994); D. Pines, Z. Phys. B to be published;
D.Pines, cond-mat/9707267.
\bibitem{kampf}
J. Altmann,W. Brening and A.P. Kampf,cond-mat/9707267.
\bibitem{timusk} C.C. Holmes {\it et al.} Phys. Rev. Lett. {\bf 71},
1645 (1993); A.V. Puchkov, D.N. Basov, and T. Timusk,
J. Phys.: Condens. Matter {\bf 8}, 10049 (1996).
\bibitem{hayden} T.E. Mason {\it et al.} Phys. Rev. Lett. {\bf 77},
1604 (1996).
\bibitem{ding} H. Ding {\it et al.}, Science {\bf 382}, 51 (1996).
\bibitem{okanderson} O.K. Andersen, O. Jepsen, A.I. Liechtenstein and
I.I. Mazin, Phys. Rev. B {\bf 49}, 4145 (1994).
\bibitem{newns} D.M. Newns, C.C. Tsuei and P.C. Pattanik, Phys. Rev. B 
{\bf 52}, 13611 (1995).
\bibitem{gyorffy} B.L. Gy\"orffy, Z. Szotek, W.M. Temmerman, O.K. Andersen
and O. Jepsen, unpublished.
\bibitem{paderef} H.J.Vidberg and J.W.Serene, J.Low Temp. Phys. {\bf
29}, 179 (1977).
\bibitem{schrieffer}
Z.-X. Shen and J.R. Schrieffer, Phys.Rev.Lett. {\bf 78} 1771 (1997).
\bibitem{newns2}
D.M.Newns, P.C.Pattnaik, and C.C.Tsuei, {\bf 43}, 3075 (1991).



\end{references}
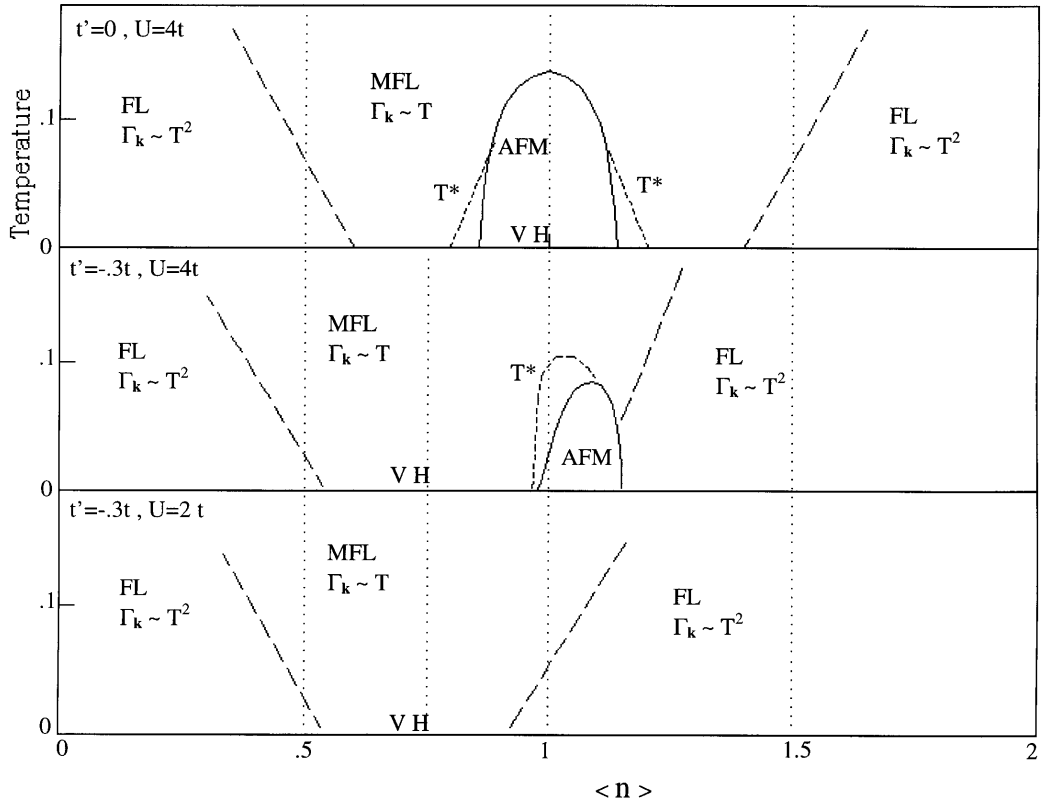
\end{document}